\documentclass{cpcauth}
\usepackage{epsfig}
\usepackage{amssymb}
\usepackage{graphicx}
\journal{Physica B}
\begin{document}
\begin{frontmatter}

\title{Magnetisation switching in a ferromagnetic Heisenberg nanoparticle with
uniaxial anisotropy: A Monte Carlo investigation}

\author{D. Ledue\thanksref{label1}},
\thanks[label1]{Corresponding author.}
\ead{denis.ledue@univ-rouen.fr}
\author{P.E. Berche},
\author{R. Patte}
\address{Groupe de Physique des Mat\'eriaux, UMR CNRS 6634,
	Universit\'e de Rouen,
	F-76801 Saint Etienne du Rouvray Cedex, France}

\begin{abstract}
We investigate the thermal activated magnetisation reversal in a single ferromagnetic
nanoparticle with uniaxial anisotropy using Monte Carlo simulations. The aim of this work is to
reproduce the reversal magnetisation by uniform rotation at very low temperature in the high
energy barrier hypothesis, that
is to realize the N\'eel-Brown model. For this purpose we have considered a simple cubic
nanoparticle where each site is occupied by a classical Heisenberg spin. The Hamiltonian is
the sum of an exchange interaction term, a single-ion anisotropy term and a Zeeman
interaction term. Our numerical data of the thermal variation of the switching field are
compared to an approximated expression and previous experimental results on Co nanoparticles.
\end{abstract}
\end{frontmatter}


\section{Introduction}
\label{sec:sec1}

The exciting field of nanostructured materials offers many challenging perspectives for
fundamental research and technological applications. Fundamentally, the
investigation of
magnetic nanoparticles has already led to a better understanding of magnetic phenomena, such as
domain structures~\cite{kittel49} and superparamagnetism~\cite{bean59}. From the
technological point of view, their potential application in magnetic recording
media~\cite{tasaki65,hayashi87} is responsible for a great interest. Indeed, sufficiently small
ferromagnetic nanoparticles are single domain~\cite{aharoni00,morrish65} and are good
candidates for high density storage. However, too small particles are superparamagnetic even at
low temperature and no information can be stored. So an optimal size has to be found for each
material.

Moreover high density storage requires a precise knowledge of the magnetisation reversal
dynamics. The investigation of the magnetisation switching of individual ferromagnetic
nanoparticles driven by an applied field at very low temperature is now possible using the
micro-SQUID technique~\cite{wernsd97a,wernsd97b,jamet01,werns01}. By waiting time, switching
field and telegraph noise measurements on
an ellipsoidal Co particle, these authors concluded that the magnetisation reversal can be
described by thermal activation over a single-energy barrier as proposed by N\'eel and
Brown~\cite{neel49,brown63}. On the other hand, because of their small size, nanoparticles are
good candidates for numerical investigations in which they can be studied with a realistic size.
Magnetisation switching in small ferromagnetic nanoparticles has been previously investigated by
Monte Carlo (MC) simulations on Ising~\cite{garcia96,richards96,richards97} or Heisenberg
models~\cite{hinzke98,hinzke99}. For highly anisotropic systems, it has been shown that the
reversal process is not uniform but rather nucleation-like. Analytical arguments and MC results
have been compared with good agreement~\cite{richards96,richards97}.

The simulation of the
uniform rotation requires models with continuous degrees of freedom, like the classical
Heisenberg model. Then, by varying the anisotropy strength or the particle size, it is possible
to investigate the crossover from uniform rotation to
nucleation~\cite{hinzke98}.
In this paper, we wish to reproduce the reversal magnetisation by uniform
rotation in a single ferromagnetic nanoparticle with uniaxial anisotropy at very low temperature.
In our case where the spins are strongly coupled and the energy barrier is large compared to $k_B T$, that is
very large metastable lifetime, the Langevin dynamics formalism~\cite{brown63} would require a huge computational
effort. Indeed, for such systems, this method is restricted to timescales of the oredr of a few
nanoseconds~\cite{nowak99}. So we have chosen Monte Carlo simulations which do not take into account the precession
process but allow to describe considerably larger timescales.
We have used the standard Metropolis algorithm.
Unlike previous works devoted to the lifetime of the metastable state~\cite{hinzke98,hinzke99},
we are interested here in the switching field which depends on the temperature but also on the
field sweeping rate. Actually, we aim to check the scaling behaviour predicted by approximated
calculations~\cite{kurkijarvi72,garg95} in the framework of the N\'eel-Brown model. Our
results will be compared to recent experimental measurements~\cite{wernsd97a,wernsd97b} which
are in good agreement with this scaling behaviour. We also wish to precise the limits
of validity of this approximated expression in terms of temperature and field sweeping rate.
The study is carried out for several particle sizes and anisotropy constant values.
In Section~\ref{sec:sec2}, we briefly remind the main predictions of the N\'eel-Brown theory and
also derive results concerning the switching field. The model and the simulation technique are
presented in Section~\ref{sec:sec3}. The results are discussed in Section~\ref{sec:sec4} and a
conclusion is given in Section~\ref{sec:sec5}.

\section{Magnetisation switching of a single nanoparticle}
\label{sec:sec2}

The stable configurations of a ferromagnetic particle of radius $R$ depend on the two ratios $R/\lambda$
and $R/ \delta$ where $\lambda\sim\sqrt{J}/M_s$ is the exchange length and $\delta\sim\sqrt{J/D}$ is the domain
wall width ($J$, $M_s$ and $D$ are, respectively, the exchange interaction, the spontaneous magnetisation and
the uniaxial anisotropy constant per atom). More precisely, the nanoparticle is single domain (with all spins aligned)
if its radius is smaller than a critical value~\cite{aharoni00}.
Such a nanoparticle exhibits
two ground states of opposite magnetisation along the easy axis, for example the $x$ axis. Under an applied field $\bf{H}$ in the
opposite direction of the particle's magnetisation, the magnetisation is expected to reverse by uniform rotation, that is,
the spins remain parallel during the reversal. According to the Stoner-Wohlfarth model~\cite{stoner48} and its generalisation
~\cite{victora89} the
energy barrier which is only
due to the anisotropy of the system can be written, at $0\ K$, as
\begin{equation}
\Delta E(H)=\Delta E_0\left (1-{H\over H_{sw}^0}\right )^\alpha
\label{eq.1}
\end{equation}
where $\Delta E_0$, the anisotropy energy barrier in zero field of a $N$-spin particle, is given
by
\begin{equation}
\Delta E_0 = NDS^2
\label{eq.2}
\end{equation}
($S$ is the spin value). The
exponent $\alpha$ depends on the direction of the applied field: $\alpha$
is equal to $2$ when the field is along the anisotropy axis or perpendicular to it and
$\alpha\approx 3/2$ else~\cite{wernsd97a}.
$H_{sw}^0$, the field
for which the energy barrier vanishes, is the switching field at $0\ K$. Assuming that the
intensity of the atomic moment is related to the spin
value by $m=g \mu_B S$ leads to
\begin{equation}
H_{sw}^0={2DS\over g\mu_0\mu_B}.
\label{eq.3}
\end{equation}
If $k_B T<< \Delta E(H)$, only the up and down magnetized states along the $x$ axis
can be observed and if $k_B T<< J$, Eqs. (\ref{eq.1}),
(\ref{eq.2}) and (\ref{eq.3}) remain roughly valid.
From the N\'eel-Brown theory~\cite{neel49,brown63},
the lifetime of the metastable state can be expressed by a thermally activated expression
\begin{equation}
\tau (T,H)=\tau_0\exp (\Delta E(H)/k_B T)
\label{eq.4}
\end{equation}
where the prefactor $\tau_0$ is usually supposed constant
for simplicity~\cite{wernsd97a}. Actually, it is the transfer of the thermal energy of
the lattice to the system of coupled spins which is responsible for the
rotation of the magnetic moments through the equatorial $y-z$ plane. So
$\tau_0$ depends on the temperature but also on the applied field~\cite{neel49,brown63,klik90,hanggi90}. 

The magnetisation reversal can be investigated by increasing the applied field over time and
measuring the switching field. Since the probability $P(t)$ that the magnetisation has not
switched after a time $t$ decreases as the temperature increases ($P(t) = \exp[-t/\tau (T,H)]$),
the switching field decreases as the temperature increases. On the other hand, the switching
field increases with the field sweeping rate $v={dH\over dt}$. Since the magnetisation reversal
by thermal activation is a stochastic process, the covered trajectory in the phase space is
different from an experiment to another, so the switching field is varying and one has to deal
with a switching field distribution. For small enough field sweeping
rates, this distribution is given by~\cite{kurkijarvi72}
\begin{equation}
p(T,H)={1\over\tau (T,H)v}\exp\left [-\int_0^{H}{dH'\over\tau(T,H')v'}\right ].
\label{eq.5}
\end{equation}
It is the product of the probability ${1\over\tau(T,H)v}$ that the reversal occurs in the
interval $[H,H+dH]$ by the probability that the reversal has not yet occured. From the Taylor's
development of $p(H)$ around its maximum, one can deduce the mean switching field by
\begin{equation}
H_{sw}(T,v)\approx H_{sw}^0\left [1-\left ({k_B T\over\Delta E_0}\ln\left ({cT\over v
\varepsilon^{\alpha-1}}\right )\right )^{1/\alpha}\right ]
\label{eq.6}
\end{equation}
where $c=k_B H_{sw}^0/(\alpha\tau_0\Delta E_0)$ and
$\varepsilon=1-H_{sw}/H_{sw}^0$~\cite{kurkijarvi72,garg95}. Then, assuming that the standard
deviation is roughly equal to the half-width of the distribution, one
obtains~\cite{kurkijarvi72,garg95}
\begin{equation}
\sigma (T,v)\approx {H_{sw}^0\over\alpha}\left ({k_B T\over\Delta E_0}\right )^{1/\alpha}
\left [\ln\left ({cT\over v\varepsilon^{\alpha-1}}\right )\right ]^{(1-\alpha)/\alpha}.
\label{eq.7}
\end{equation}
This quantity increases with temperature and with the field sweeping rate.

\section{Model and simulation technique}
\label{sec:sec3}

\subsection{Model}

Our model consists of a simple cubic lattice included in a sphere of radius $R$ with free
boundary conditions. Each site is occupied by a classical Heisenberg
spin ${\bf S}_i=(S_i^x,S_i^y,S_i^z)$  of modulus $S$ in order to reproduce uniform rotation and
to investigate the effect of the anisotropy strength. The hamiltonian $H$ is given by
\begin{equation}
H=-J\sum_{<i,j>}{\bf S}_i.{\bf S}_j-D\sum_i (S_i^x)^2-{\bf H}.\sum_i{\bf S}_i
\label{eq.8}
\end{equation}
where $J>0$ is the ferromagnetic exchange interaction limited to the nearest neighbors and
$D > 0$ is the anisotropy constant on each site. The first term favors alignment of all spins
and the second one favors alignment along the $x$ axis (up or down). The last sum represents the
interaction with the applied field. Here, we have considered an applied field along the $x$ axis:
${\bf H} = -H {\bf {\hat e}}_x$ where ${\bf {\hat e}}_x$ is the unit vector and $H > 0$. Then,
the ground
state corresponds to all spins being antiparallel to the unit vector ${\bf {\hat e}}_x$ whereas
the metastable state corresponds to all spins pointing up in the $x$ direction. We have to
mention that we have taken here the same exchange interactions, anisotropy constants and anisotropy axis for the
core and the surface of the nanoparticle although surface anisotropy is usually considered as radial and stronger. Since our
calculations are restricted to very small
nanoparticles, much smaller than the critical size to be single-domain in the remanent state~\cite{aharoni88}, we have
neglected dipolar interactions. The magnetisation per spin of the particle is defined by

$${\bf M} = {1\over N}\sum_{i=1}^N{\bf S}_i.$$ 

\subsection{Simulation technique}

The numerical procedure is the importance-sampling MC
method~\cite{heermann90,binder90,mackeown97}. In our simulations, we have used the
standard Metropolis algorithm~\cite{metropolis53}. Each simulation is
performed at a given temperature $T$. The particle is initially magnetized in the $x$
direction by
applying a strong field ${\bf H_0} = H_0 {\bf {\hat e}}_x$ ($H_0 > 0$) during $n$ MC steps (a MC
step (MCS) corresponds to the scan of all spins once, trying a rotation for each). Then, the
antiparallel applied field is ramping by small jumps $\delta H$: ${\bf H}_{\it p}= -p\ \delta H\ 
{\bf {\hat e}}_x$.
Since the field is kept constant during $n$ MCS, the field sweeping rate is given by
$v=\delta H/ n$.
The simulation is stopped when the magnetisation reversal is observed (the arbitrary chosen
criterion is  $M_x < -0.8 S$) and the
switching field value is stored. For each temperature and each value of the field sweeping rate,
a relatively large number of simulations $n_s$ has to be performed to get a reliable estimate of
the mean switching field $H_{sw}(T,v)$. This has been done using a parallel version of the code.
The error for $H_{sw}(T,v)$ will be given by
$\sigma(T,v)/\sqrt{n_s}+\delta H/n_s$ where $\sigma (T,v)$ is the numerical estimate of
the standard deviation of the switching field distribution.

Starting from the initial state magnetized in the $x$ direction, a uniform
rotation of the magnetisation of an angle $\theta$ produces, using Eq. (\ref{eq.8}), an energy
variation
$$\delta E=NS (1-\cos \theta)[DS (1+\cos \theta)-H]$$
where there is no exchange energy. Simulating this mechanism by a single spin rotation (SSR)
algorithm implies an increase of the exchange energy at each individual rotation of $\theta$.
The corresponding energy variation of the system is
$$\delta E_{{\rm SSR}}=S (1-\cos \theta)\left [zJS+DS (1+\cos \theta)-H\right ]$$
where $z$ is the coordination number. Since the acceptance of the SSR requires
$\delta E_{{\rm SSR}}\leq 0$ at $0\ K$, no individual rotation can occur at $0\ K$ if
$H <H_{{\rm SSR}}^c=zJS+DS (1+\cos \theta)$ and, consequently, no uniform rotation of
all spins. This clearly means that the algorithm fails at $0\ K$ since it gives a switching field
value larger than the field which produces the flip of a surface spin ($z=1$), that is $H=JS$
which has nothing to do with the expected value $H_{sw}^0$.
At very low temperature and small applied field as in our simulations (that is $\delta
E_{{\rm SSR}}> 0$), SSR has a very small occurence probability and uniform rotation itself
can require a very large number of MCS. So, the switching field at very low temperature will be
overestimated except for small enough field sweeping rates. There is a temperature range and field
sweeping rate range for which the algorithm dynamics is suitable for the investigation of the
uniform rotation.

\section{Numerical results}
\label{sec:sec4}

The spin value has been fixed to $S = 1$ for all nanoparticles. The physical parameters $N$, $J$ and $D$ have to be
chosen carefully in order to compare our results with previous experimental ones with a reasonable
computational effort, that is small sizes. Actually, the magnetisation reversal process depends, as the stable 
states, on the two ratios $R/ \lambda$ and $R/ \delta$. In their experiments, W. Wernsdorfer {\it et al.}~\cite{wernsd97a}
have investigated a single domain Co nanoparticle with a radius of $(12.5\pm 2.5)$ nm which yields $R/ \lambda\simeq 1.5$ and
$R/\delta\simeq 0.7$.
Here, we have considered a $33-$spin nanoparticle $(R=2a,$ $a$ is the lattice parameter) with $J=k_B$ and $D=0.1\ J$ so
$R/\lambda\simeq 1$ and
$R/\delta\simeq 0.3$. These values are of the same magnitude as the experimental ones. Actually, in our model, we have
underestimated the exchange interaction of a factor $250$ in comparison to Co which decreases the two
characteristic lengths $\lambda$ and $\delta$ and allow us to investigate very small nanoparticles.
The choice of $J$ has only effect on
the simulated temperatures since we have imposed the low temperature condition $k_B T<<J$ ($0.01 \leq k_B T/J \leq 0.1$).
It has to be mentioned that the quality factor $Q=(\lambda/\delta)^2$ which defines the hardness of the ferromagnetic
material is equal to 0.09 in our model, of the same magnitude as the Co value 0.2.
The mean switching field, at a given temperature and for a given field
sweeping rate, was obtained by averaging over $100$ different realisations.

\begin{figure}[t]
	\epsfysize=6cm
	\begin{center}
	\mbox{\epsfbox{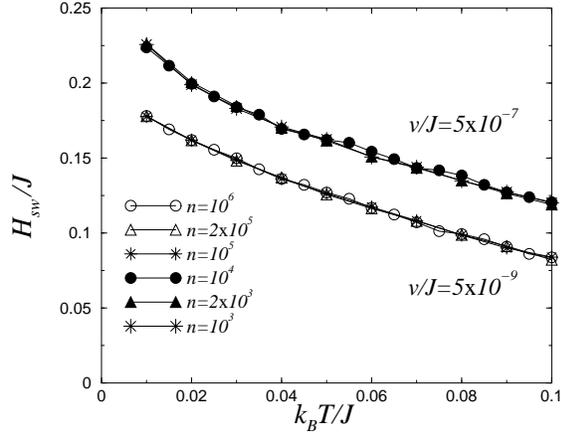}}
	\end{center}\vskip 0cm
	\caption{Thermal variation of the mean switching field for 
$\delta H/n=5\times 10^{-7}\ J$ and $5\times 10^{-9}\ J$
for three different values of $(n,\delta H)$ ($N=33$ spins and $D=0.1\ J$). The error bars are
smaller than the symbols.}
	\label{fig1}
\end{figure}

Firstly, we checked that our definition of the field sweeping rate $v=\delta H/n$  is
reliable; then, we have determined the thermal variation of the switching field
of a given particle with different values of $\delta H$ and $n$ such that
$\delta H/n$ is kept constant. In Fig.~\ref{fig1}, we have plotted the two cases
$\delta H/n=5\times 10^{-7}\ J$ and $5\times 10^{-9}\ J$. It can be seen in both cases that the
three curves corresponding to different numbers of MCS are almost identical indicating that
$\delta H/n$ is a reliable definition for the
field sweeping rate. The fluctuations of the data are smaller for $v=5\times 10^{-9}\ J$.
In the following, we
have fixed $\delta H = 5\times 10^{-3}\ J$ and only $n$ is varying (Table~\ref{Tab1}).

      \begin{table}
        \caption{Field sweeping rate versus the number of MCS for $\delta H=5\times 10^{-3}\ J$.
        \label{Tab1}}
       \vspace*{0.1cm}
        \begin{tabular}{|l|l|l|l|l|l|l|l|l|l|}  \hline
        $n\ (\times 10^3)$ & $5$  & $10$ & $50$ & $100$ & $500$ & $1000$ & $5000$ & $10000$ & $100000$ \\  \hline
        $v/J\ (\times 10^{-6})$ & $1$ & $0.5$ & $0.1$ & $0.05$ & $0.01$ &
$0.005$ & $0.001$ & $0.0005$ & $0.00005$ \\  \hline
        \end{tabular}
        \end{table}

\subsection{Temperature and field sweeping rate dependence of the reversal field}

\begin{figure}[t]
	\epsfysize=5.8cm
	\begin{center}
	\mbox{\epsfbox{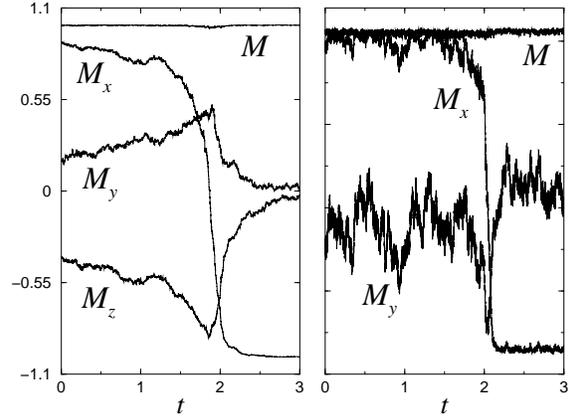}}
	\end{center}\vskip 0cm
	\caption{Time dependence in $10^4$ MCS of the modulus and the components
of the magnetisation per spin at $k_B T/J = 0.01$ (left side) and $k_BT/J = 0.1$ (right side) 
($N = 33$ and $D = 0.1\ J$). For reason of clarity, the $M_z$ component is not shown on the
right side.
Note that we have plotted the time evolution for only three values of the field.}
	\label{fig2}  
\end{figure}

We present here our numerical results concerning a nanoparticle of $N=33$ spin with an
anisotropy constant $D=0.1\ J$.
In Fig.~\ref{fig2}, we have plotted the time evolution of the modulus and the components
of the magnetisation per spin for $k_B T/J =0.01$ and $0.1$.
For $k_B T/J = 0.01$, the thermal fluctuations of the modulus
are very small and the uniform rotation is observed (the modulus is roughly constant during the
reversal). The reversal duration is about $4\times 10^4$ MCS which means that the reversal
dynamics is slow. For $k_B T/J = 0.1$, the reversal is still uniform but
thermal fluctuations are visible. Many reversal attempts can be seen before the
reversal itself whose duration is significantly reduced
to about $4\times 10^3$ MCS. We can conclude that these parameters allow to be in the uniform rotation
regime unlike the case of larger anisotropy constant, as it will be seen later.

\begin{figure}[t]
	\epsfysize=6.4cm
	\begin{center}
	\mbox{\epsfbox{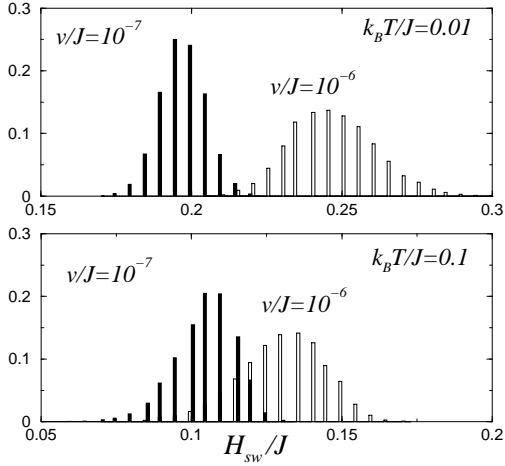}}
	\end{center}\vskip 0cm
	\caption{Probability distribution of the switching field at $k_B T/J=0.01$ and
$0.1$ for $v/J=10^{-6}$ and $10^{-7}$ ($N=33$ and $D=0.1\ J$). Each histogram is obtained
from $5000$ samples.}
	\label{fig3}  
\end{figure}

The probability distribution of the switching field for $5000$ samples is plotted in
Fig.~\ref{fig3} at $k_B T/J=0.01$ and $0.1$ for two different field sweeping rates
$v/J=10^{-6}$ and $10^{-7}$. We clearly see the gaussian shape of the distribution whose width
proportional to $\sigma$ increases with $v$ and $T$ in agreement with Eq. (\ref{eq.7}).

The field sweeping rate dependence of the mean switching field for several temperatures is shown
in Fig.~\ref{fig4}. An almost
logarithmic dependence at small sweeping rate can be observed as in Ref.~\ref{wernsd97a} whose
experimental results are shown for comparison. In these experiments the mean switching field has been measured
over several hundred cycles.

\begin{figure}
	\epsfysize=6cm
	\begin{center}
	\mbox{\epsfbox{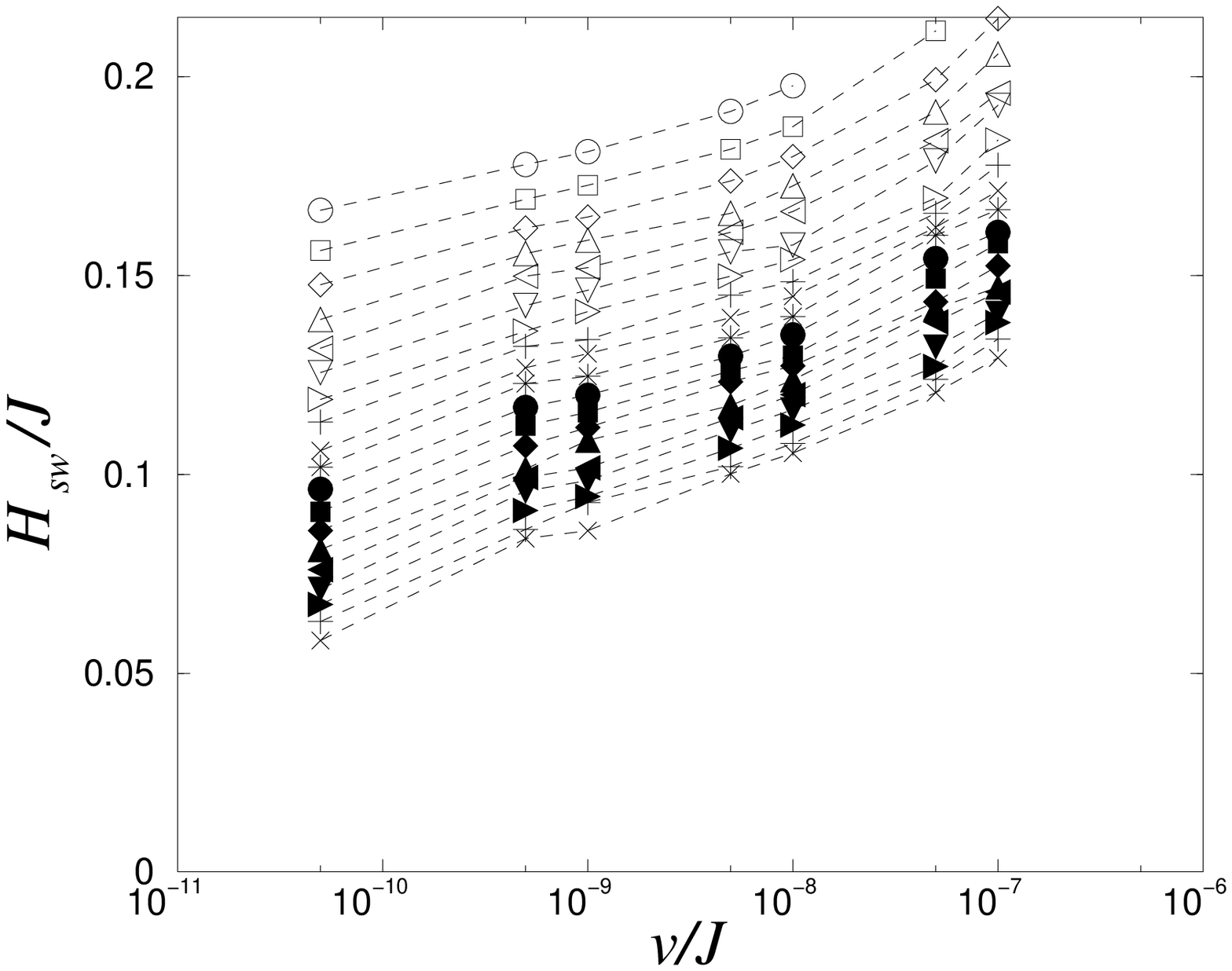}}
	\end{center}
\vspace{-3.5cm}
\end{figure}
\begin{figure}
	\epsfysize=2.3cm
	\epsfxsize=2.5cm
        \hspace{4.5cm}
	\mbox{\epsfbox{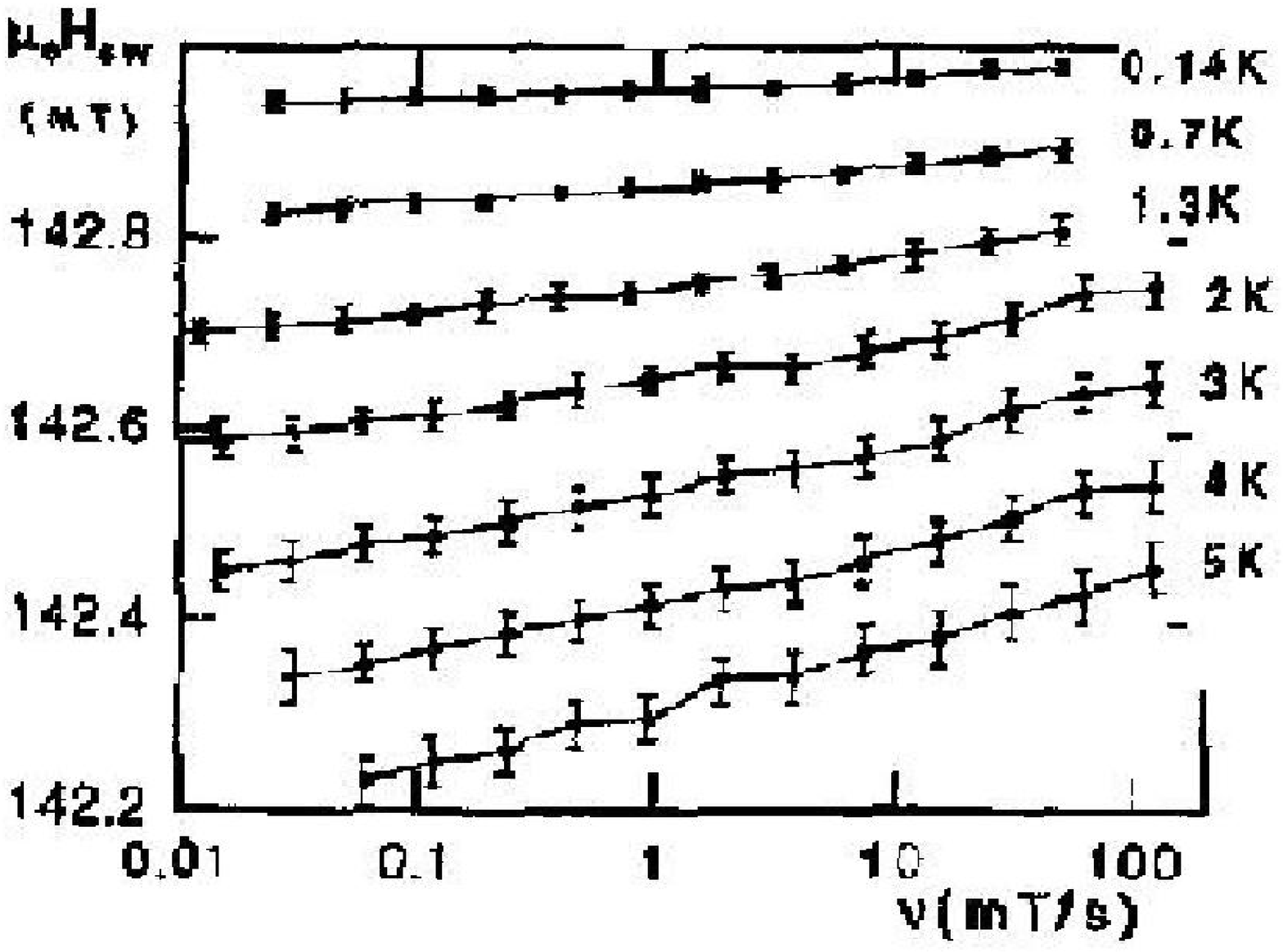}}
\vspace{0.5cm}
	\caption{Field sweeping rate dependence of the mean switching field of a $33$-spin
particle with $D = 0.1\ J$ for different temperatures $0.01 \leq k_B T/J \leq 0.1$
(increasing temperature from the top to the bottom). Insert: Experimental results of
Wernsdorfer et al~\cite{wernsd97a}.}
	\label{fig4}
\end{figure}

\begin{figure}
	\epsfysize=6.2cm
	\begin{center}
	\mbox{\epsfbox{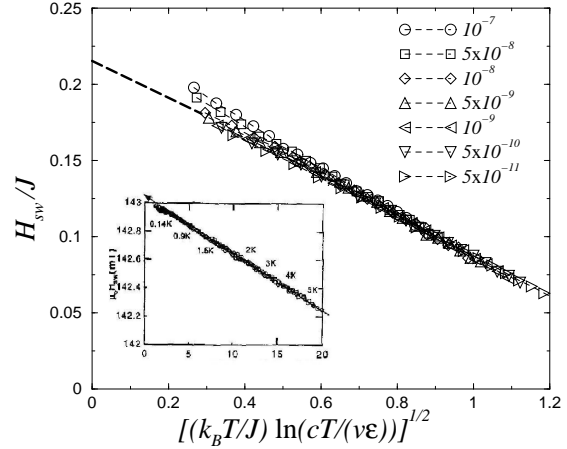}}
	\end{center}
	\caption{Scaling plot of the mean switching field values of a $33$-spin particle with
$D = 0.1\ J$ for $5\times 10^{-11}\leq v/J\leq 10^{-7}$ and $0.01 \leq k_B T/J \leq 0.1$.
The dashed line shows the mean linear fit. Insert: experimental results of
Wernsdorfer et al~\cite{wernsd97a}.}
	\label{fig5}
\vspace{-6.05cm}
\end{figure}

To check the validity  of Eq. (\ref{eq.6}), one can
plot the mean switching field values versus $\left [k_B T\ln\left ({cT\over v
\varepsilon}\right )\right ]^{1/2}$ where $\varepsilon$ depends on $H_{sw}$. If the scaling
behaviour is satisfied, all data points
should collapse on a single straight line by choosing correctly the constants $H_{sw}^0$ and $c$.
Then, a self-consistent fit is obtained when the
vertical axis intercept is the chosen value $H_{sw}^0$. From Eq. (\ref{eq.6})~, the slope is
$-H_{sw}^0 /(\Delta E_0)^{1/2}$ from which it is possible to extract an estimate of $\Delta E_0$.
Putting back $H_{sw}^0$ and $\Delta E_0$ in the expression of $c$ provides a numerical value
for $\tau_0$. Such a fit is shown in Fig.~\ref{fig5}.

\begin{figure}[t]
	\epsfxsize=3cm
	\epsfysize=2.4cm
         \hspace{1.3cm}
	\mbox{\epsfbox{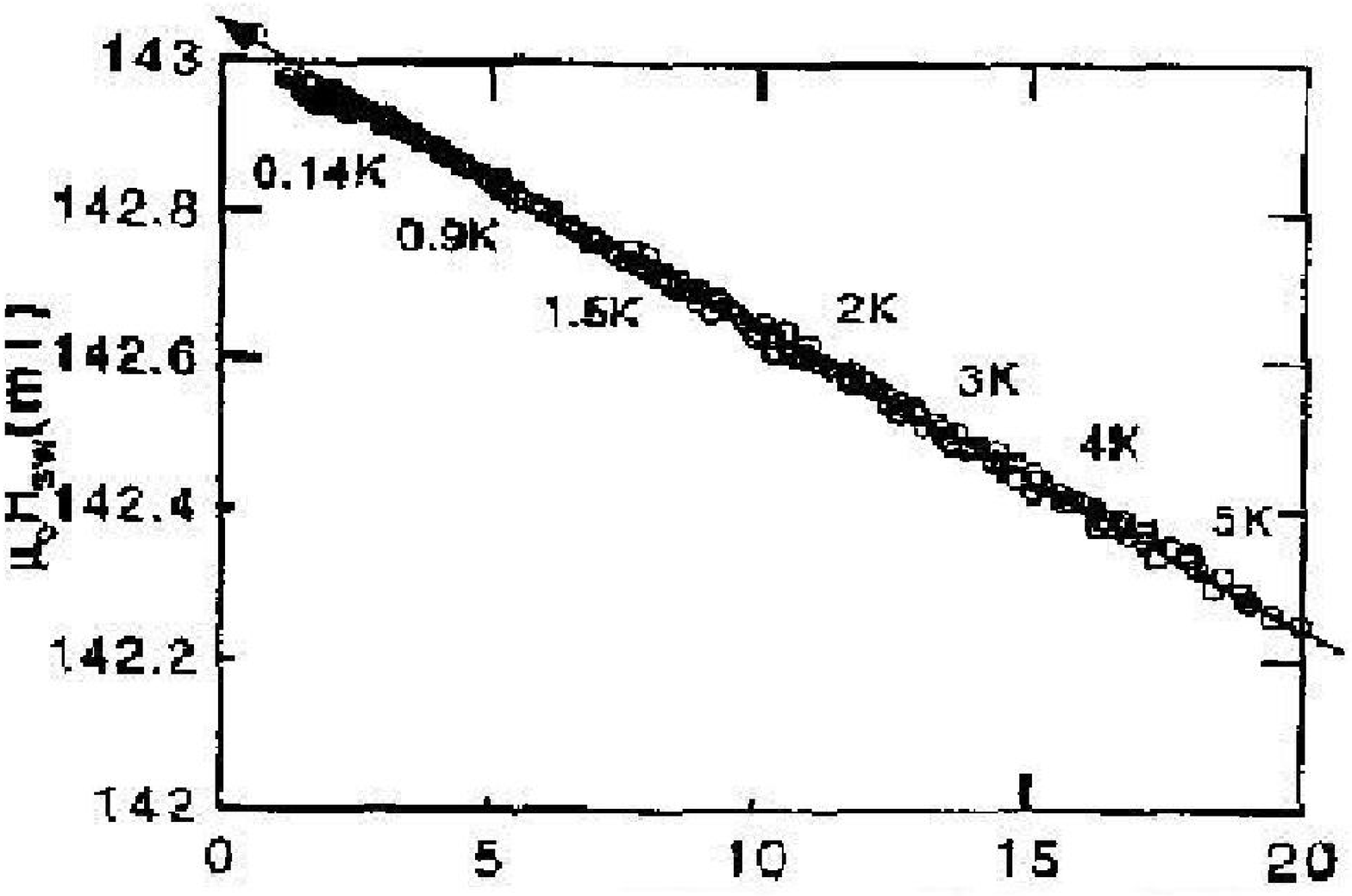}}
\vspace{3.4cm}
\end{figure}

Only the curves corresponding to the smallest field sweeping rates ($v/J\leq 10^{-8}$) collapse
on a single straight
line. Numerical data associated to higher sweeping rates move away from the scaling behaviour as
the temperature decreases.
This deviation
is attributed to the inefficiency of the SSR procedure at very low temperature
(see Section~\ref{sec:sec3}.2) because our numerical switching field values can be larger than
$H_{sw}^0$ which is not physical.
No deviation due to
thermal fluctuations has been noticed even when $k_B T/J\rightarrow 0.1$. The results of the best fit
obtained with $c/k_B=10^{-3}$ and $H_{sw}^0=0.215\ J$ are reported in TABLE~\ref{Tab2}. The vertical
axis intercept is consistent with the value used to perform the fit but is slightly higher than
the expected one $H_{sw}^0\simeq 0.2\ J$ (Eq. (\ref{eq.3}) with $g\mu_0\mu_B =1$). Our
estimate of the energy barrier in zero field $2.8\ J$ is
lower than the expected value $\Delta E_0\simeq 3.3\ J$ (Eq. (\ref{eq.2})).

      \begin{table}
        \caption{Parameters of the fits and estimates of the switching field at $0\ K$ and
the energy barrier in zero field for two particle sizes and $D = 0.1\ J$. The axis
intercept corresponds to $H_{sw}^0 /J$ and the slope to $-(H_{sw}^0/J)/(\Delta E_0/J)^{1/2}$.
        \label{Tab2}}
       \vspace*{0.1cm}
        \begin{tabular}{|l|l|l|}   \hline 
        $N$ & $33$ & $123$ \\ \hline 
        $H_{sw}^0/J$ & $0.215$ & $0.215$ \\ 
        $c/k_B$  & $10^{-3}$ & $10^{-4}$ \\
        axis intercept & $0.219$ & $0.215$ \\
        slope  & $-0.132$ & $-0.080$ \\  
        $\Delta E_0/J$  & $2.8$ & $7.2$ \\
        $\tau_0$  & $40$ & $150$  \\ \hline
        \end{tabular}
        \end{table}

\subsection{Influence of the particle size on the reversal field}

Since the magnetisation reversal mechanism is very sensitive to the system size, we have studied
the influence of the number of spins on the mean switching field variation in the case
$D=0.1\ J$.

For $N = 7$ ($R=a$), the time dependence of the modulus and the components of the magnetisation exhibits
no significant difference from the case $N = 33$ except an increase of the thermal fluctuations.
Indeed, for a given temperature, thermal fluctuations increase as the size of the system
decreases. A consequence is that the reversal duration is reduced to about $6\times 10^3$ MCS and
$10^3$ MCS at $k_B T/J =0.01$ and $0.1$ respectively. For $N=123$ ($R=3a$), the two ratios $R/\lambda\simeq 1.5$ and
$R/\delta\simeq 0.4$ are closer to the experimental values; in that case, we have observed again that
the reversal is uniform which agrees with experimental results~\cite{wernsd97a}.

\begin{figure}[t]
	\epsfysize=6cm
	\begin{center}
	\mbox{\epsfbox{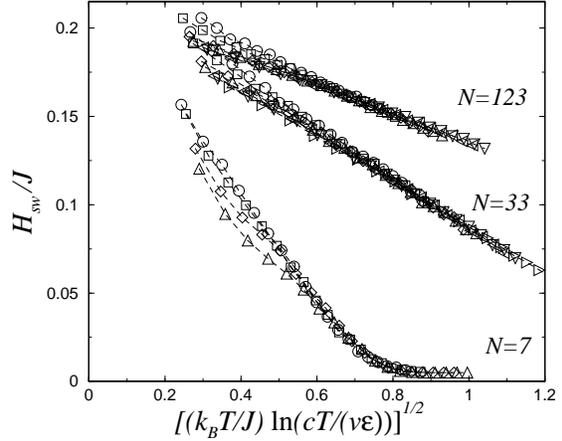}}
	\end{center}\vskip 0cm
	\caption{Scaling plot of the mean switching field values of a $7$-spin, a $33$-spin
and a $123$-spin particle with $D = 0.1\ J$ and $0.01 \leq k_B T/J \leq 0.1$. The values of the
field sweeping rate are $5\times 10^{-9}\leq v/J \leq 10^{-7}$ for $N=7$, 
$5\times 10^{-11}\leq v/J \leq 10^{-7}$ for $N=33$ and $5\times 10^{-10}\leq v/J \leq 10^{-7}$ for $N=123$.}
	\label{fig6}  
\end{figure}

The plot of the mean switching field values versus $\left [k_B T\ln\left ({cT\over v
\varepsilon}\right )\right ]^{1/2}$ is drawn in Fig.~\ref{fig6} for $N=7$, $33$ and $123$.

For $N = 123$, using the same procedure as in the previous section, we observed similar results
as for $N = 33$. A self-consistent fit can be obtained with the same value of
$H_{sw}^0$ as for $N=33$ in agreement with the fact that $H_{sw}^0$ should not depend on the
particle size (TABLE~\ref{Tab2}). The extracted energy barrier in zero field is significantly
lower than the expected value $\Delta E_0\simeq 12.3\ J$ again.

In the case $N=7$, the curvature observed as the temperature increases corresponds to the
transition towards the superparamagnetic regime. Indeed, the particle is still strongly magnetized
at these temperatures (the time average of the modulus of $\bf{M}$ at $k_B T/J=0.1$ is
$0.934 S$), so the zero mean switching field (actually $\delta H$) obtained for the smallest field
sweeping rates means that the magnetisation spontaneously reverses in zero field. At low temperature,
the data do not lie on a single straight line. The deviation observed for $v/J=5\times 10^{-9}$ and
$10^{-8}$ is attributed to unlikely single flips of spins near the particle boundary (the coordinence
number is only $1$ for $6$ spins) which induce the magnetisation reversal earlier as expected.
For this reason, we did not perform simulations with smaller field sweeping rates.
This deviation has not been
observed for $N=33$ and $123$ because surface effects are much less pronounced.

\subsection{Influence of the anisotropy constant on the reversal field}

The ratio $D/J$ is also supposed to play an important role in the magnetisation reversal process.
Here, we have investigated for a $33$-spin nanoparticle the effect of increasing the
anisotropy strength ($D=0.2\ J,\ 0.3\ J$ and $J$) to evidence the
transition to a non uniform rotation regime and to study its consequence on the switching field
variation.

For $D = 0.3\ J$ ($R/\delta\simeq 0.5$), the reversal is not perfectly uniform since the jump of the modulus of the
magnetisation during the reversal is about $10\ \%$ at $k_B T/J = 0.01$ (for $D=0.2\ J$, the
jump is around
$6\ \%$). The reversal duration for $D=0.3\ J$ at
$k_B T/J = 0.01$ is about $2\times 10^4$ MCS which is of the same order of magnitude as for
$D = 0.1\ J$ and $0.2\ J$. 

\begin{figure}[t]
	\epsfysize=5.8cm
	\begin{center}
	\mbox{\epsfbox{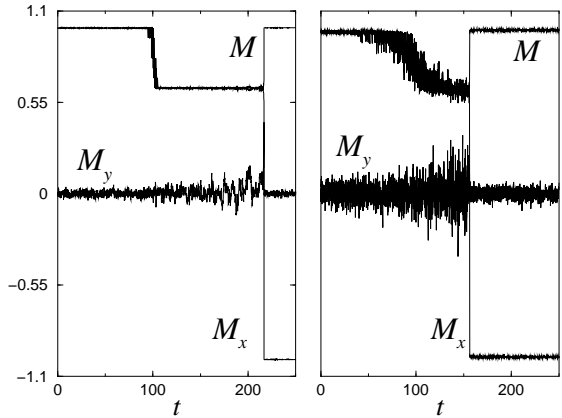}}
	\end{center}\vskip 0cm
	\caption{Time dependence in $10^4$ MCS of the modulus and the $M_x$, $M_y$ components of
the magnetisation per spin at $k_BT/J = 0.01$ (left side) and $k_B T/J=0.1$ (right side)
($N = 33$ and $D = J$).}
	\label{fig7}  
\end{figure}

For $D = J$ ($R/\delta\simeq 0.9$) the material of our model is harder than Co (the quality factor is about 0.8).
As it can be seen in Fig.~\ref{fig7}, the reversal is
clearly not uniform (the jump is about $36\ \%$). Because of strong anisotropy, the spins behave
roughly as Ising spins and the magnetisation is always along the $x$ axis except short time
scale fluctuations. This was not the case for $D=0.1\ J$ (see Fig.~\ref{fig2}). Actually the
spins near the surface of the nanoparticle firstly flip whereas the other spins with higher
coordination number flip later, that is for a larger value of the field.
At $k_B T/J=0.01$, the magnetisation reversal occurs in only two steps indicating that the
nanoparticle behaves as a core surrounded by a surface shell. At $k_B T/J=0.1$, the reversal is
more continuous due to thermal activation.

\begin{figure}[t]
	\epsfysize=6cm
	\begin{center}
	\mbox{\epsfbox{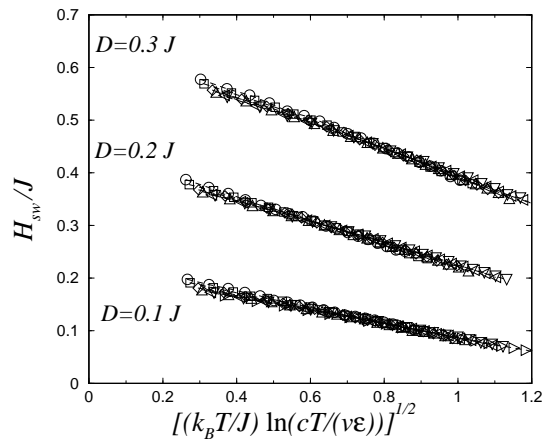}}
	\end{center}\vskip 0cm
	\caption{Scaling plot of the mean switching field values of a $33$-spin particle with 
$D = 0.1\ J$ for $5\times 10^{-11}\leq v/J \leq 10^{-7}$, $D=0.2\ J$ and $0.3\ J$ for
$5\times 10^{-10}\leq v/J \leq 10^{-7}$ and $0.01 \leq k_BT/J \leq 0.1$.}
	\label{fig8}  
\end{figure}

To check the validity of Eq. (\ref{eq.6}) for $D = 0.2\ J$ and $0.3\ J$, we have plotted the mean switching field
values using the same procedure as in the previous sections. Although the reversal is not
perfectly uniform, it is possible to put the data points corresponding to $v/J\leq 10^{-8}$ on
a single
straight line (Fig.~\ref{fig8}).
As for $D = 0.1\ J$, the scaling behaviour is satisfied for all
temperatures. 
The results of the two fits are reported in TABLE~\ref{Tab3} in comparison with the case
$D=0.1\ J$. As for $D = 0.1\ J$,
a self-consistent fit
is obtained with $H_{sw}^0$ slightly higher than the expected value but our estimates of $H_{sw}^0$
are proportional to the anisotropy constant in agreement with the Stoner-Wohlfarth model. The extracted
$\Delta E_0$ is lower than the expected value again, the increase of the misfit with
the anisotropy constant being attributed to the evolution towards a non-uniform rotation.

      \begin{table}
        \caption{Parameters of the fits and estimates of the switching field at $0\ K$ and 
the energy barrier in zero field for a $33$-spin particle with different values of the
anisotropy constant.}
        \label{Tab3}
       \vspace*{0.1cm}
        \begin{tabular}{|l|l|l|l|l|l|l|l|} \hline
        $D/J$ & $0.1$ & $0.2$ & $0.3$ \\ \hline
        $H_{sw}^0/J$  & $0.215$ & $0.430$ & $0.645$ \\
        $c/k_B$ & $10^{-3}$ & $10^{-3}$ & $10^{-2}$ \\
        axis intercept & $0.219$ & $0.432$ & $0.646$ \\
        slope & $-0.132$ & $-0.207$ & $-0.252$ \\
        $\Delta E_0/J$ & $2.8$ & $4.3$ & $6.6$ \\
        $\tau_0$ & $40$ & $50$ & $5$  \\ \hline
        \end{tabular}
        \end{table}

\begin{figure}[t]
	\epsfysize=6cm
	\begin{center}
	\mbox{\epsfbox{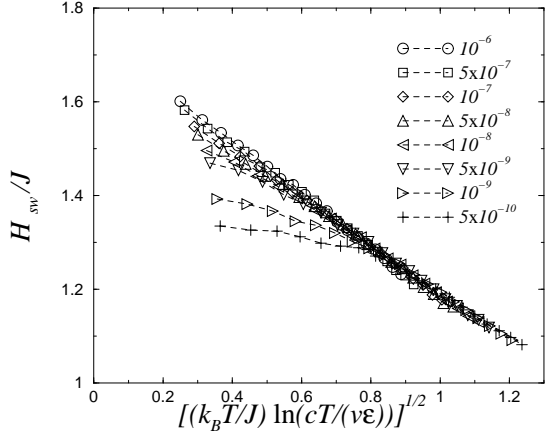}}
	\end{center}\vskip 0cm
	\caption{Scaling plot of the mean switching field values of a $33$-spin particle
with $D = J$ for $5\times 10^{-10}\leq v/J \leq 10^{-6}$ and $0.01 \leq k_B T/J \leq 0.1$.}
	\label{fig9}  
\end{figure}

For $D = J$, it is not possible to put the low temperature data points on a single straight line. The best fit
is shown in Fig.~\ref{fig9}. Note that the switching field values are significantly lower than that predicted
in the case of uniform rotation ($H_{sw}^0 /J=2$).
Actually, at very low temperature, 
only spin flips can occur, so the switching field does not depend anymore on the anisotropy
constant but rather on the exchange interaction and the coordination number. We expect that
the numerical estimate of the switching field at very low temperature would have not been changed increasing
the value of the anisotropy constant.

\section{Conclusion}
\label{sec:sec5}

Using a MC single spin rotation algorithm, we have realized the N\'eel-Brown model of the
magnetisation reversal for nanoparticles with realistic values of $R/\lambda$ and
$R/\delta$. By choosing the temperature and the field
sweeping rate appropriately, it has been possible to check the validity of the approximated
expression of the mean switching field. For reasonable
anisotropy ($D\leq 0.3\ J$), the scaling behaviour is satisfied at very small field sweeping
rates ($v/J\leq 10^{-8}$) for all temperatures studied except the case $N=7$ and $D=0.1\ J$
for which the superparamagnetic regime can be observed. On the other hand, for high anisotropic nanoparticles
($D = J$), the magnetisation reversal is not uniform and the
mean switching field values can not be fitted anymore following the proposed
approximated expression.

Let us briefly remind that our
estimates of $H_{sw}^0$  are independent of the size and vary linearly with the anisotropy
constant, as expected. The discrepancy between the estimate of $\Delta E_0$
and its expected value in all cases might be due to the spin rotation procedure.

In a near future, it is planned to investigate the effect of physical parameters, such as surface
anisotropy and dipolar
interactions, in relation with the shape of the particle. However, one has to be aware of the
very large increase of CPU time by taking into account the dipolar interactions.

\vspace{0.5cm}
\noindent {\bf Acknowledgments}
\vspace{0.5cm}

This work was supported by the computer-time grants 2000006 of the Centre 
de  Ressources Informatiques de Haute Normandie (CRIHAN) and C20020922345 of the Centre
Informatique National de l'Enseignement Sup\'erieur (CINES). We are indebted to W. Wernsdorfer
and N. Lecoq for interesting discussions.

\def\paper#1#2#3#4#5{#1, #3 {\bf #4}, \rm #5 (#2).}
\def\papers#1#2#3#4#5{#1, #3 {\bf #4}, \rm #5 (#2)}
 	\def\PRB{Phys. Rev. B}
 	\def\PRE{Phys. Rev. E}
 	\def\PRL{Phys. Rev. Lett.}
	\def\JPA{J. Phys. A: Math. Gen.}


\end{document}